\documentstyle[12pt]{elsart}
\begin{document}

\begin{frontmatter}
\title{One-Photon Transitions between Heavy Baryons
in a Relativistic Three-Quark Model}
\author{M.\ A.\ Ivanov}
\address{Bogoliubov Laboratory of Theoretical Physics, Joint Institute
for Nuclear Research, 141980 Dubna, Russia}
\author{J.\ G.\ K\"{o}rner}
\address{Johannes Gutenberg-Universit\"{a}t, Institut f\"{u}r Physik,
Staudinger Weg 7, D-55099 Mainz, Germany}
\author{V.\ E.\ Lyubovitskij}
\address{Bogoliubov Laboratory of Theoretical Physics, Joint Institute
for Nuclear Research, 141980 Dubna, Russia
and Department of Physics, Tomsk State University, 634050 Tomsk, Russia}
\maketitle

\begin{abstract}
We study one-photon transitions between heavy baryon states in the
framework of a relativistic three-quark model. We calculate the one-photon
transition rates for ground-state to ground-state transitions and for
some specific excited state to ground-state transitions.
Our rate predictions for the most important transitions are:
$\Gamma(\Sigma^{+}_{c}\rightarrow \Lambda_c^+\gamma)= 60.7\pm 1.5 $ KeV,
$\Gamma(\Xi^{'+}_{c}\rightarrow \Xi_c^+\gamma) = 12.7\pm 1.5$ KeV,
$\Gamma(\Lambda_{c1}(2593)\rightarrow\Lambda_c^+\gamma)=104.3\pm 1.3$ KeV.

\end{abstract}
\end{frontmatter}



\newpage
During the last few years there has been significant progress in the
experimetal study of the spectroscopy of ground state and excited state
charm baryons and their strong and electromagnetic decays~\cite{PDG,CLEO_98}.
The one-photon transitions $\Lambda_c^{*+}(2593)\to\Lambda_c\gamma$ and
$\Lambda_c^{*+}(2625)\to\Lambda_c\gamma$ were searched for by the
CLEO Collaboration but were not seen\cite{CLEO}. The CLEO Collaboration
determined upper limits for the branching ratios
$$
B(\Lambda_c^{*+}(2593)\to\Lambda_c^+\gamma)/
B(\Lambda_c^{*+}(2593)\to\Lambda_c^+\pi^+\pi^-)<0.98,$$
$$
B(\Lambda_c^{*+}(2625)\to\Lambda_c^+\gamma)/
B(\Lambda_c^{*+}(2625)\to\Lambda_c^+\pi^+\pi^-)<0.528
\footnote{In the following we use the notation $\Lambda_{c1}$ and
$\Lambda_{c1}^{*}$ for the
excited baryon states $\Lambda_c^*(2593)$ and $\Lambda_c^*(2625)$,
respectively.}
$$
One-photon transitions between charm baryons have been analyzed before in
the leading order of the heavy quark mass expansion~\cite{Korner,Cho}, in the
nonrelativistic quark model incorparating heavy-quark symmetry~\cite{Cheng}
and in the bound state picture~\cite{Chow}. When applying heavy quark symmetry
to the one-photon transitions one makes no assumptions about the composition
of the light-side diquark states mediating the one-photon transitions. A first
rough estimate of the unknown coupling parameters entering the effective heavy
quark symmetry Lagrangian including electro-magnetism has been attempted using
simple dimensional arguments~\cite{Cho}. The light-side diquark transitions
have been calculated within the constituent quark model~\cite{Cheng} and
within a bound state model~\cite{Chow} where the heavy baryon is composed of
a heavy meson and a light baryon.
In the nonrelativistic constituent quark model one obtains for the
ground-state to ground-state transitions~\cite{Cheng}
$$\Gamma(\Sigma_c^+\to\Lambda_c^+\gamma)=93 \mbox{KeV},\,\,
\Gamma(\Xi_c^{'+}\to\Xi_c^+\gamma)=16 \mbox{KeV},\,\,
\Gamma(\Xi_c^{'0}\to\Xi_c^0\gamma)=0.3 \mbox{KeV}.$$
In the bound state picture~\cite{Chow} the
$\Lambda_{c1}\to\Lambda_c\gamma$ and $\Lambda_{c1}^*\to\Lambda_c\gamma$ decays
are severely suppressed, whereas the corresponding transitions
between bottom baryons are predicted to have significant branching ratios.

In this paper we report on the predictions of the Relativistic Three-Quark
Model~\cite{YadFiz} for the one-photon transitions between heavy baryon
states. The Relativistic Three-Quark Model was applied before
to a number of different dynamical problems involving
the properties of pions~\cite{YadFiz}, light baryons~\cite{FBS} and
heavy-light baryons~\cite{PRD97}-\cite{MZ98}. In the most recent
application the Relativistic Three-Quark Model was used to evaluate the
one-pion transition strengths between charm baryons \cite{MZ98}.

The Lagrangian describing the couplings of a heavy baryon state to its
constituent light and heavy quarks considerably simplifies in the heavy quark
limit since the heavy quark field enters as a local field and can
be factored from the nonlocal Lagrangian. One has
\begin{eqnarray}\label{HB_Lagr}
{\cal L}_{B_Q}^{\rm int}(x)&=&g_{B_Q}\bar B_Q(x)\Gamma_1 Q^a(x)
\int d^4\xi_1\int d^4\xi_2F_B(\xi_1^2+\xi_2^2)\label{L-int}\\
&\times&q^b(x+3\xi_1-\sqrt{3}\xi_2 )C\Gamma_2\lambda_{B_Q}
q^c(x+3\xi_1+\sqrt{3}\xi_2)\varepsilon^{abc}+{\rm h.c.}\nonumber\\
F_B(\xi_1^2+\xi_2^2)&=&\int\hspace*{-0.1cm}\frac{d^4k_1}{(2\pi)^4}
\int\hspace*{-0.1cm}\frac{d^4k_2}{(2\pi)^4}e^{ik_1\xi_1+ik_2\xi_2}
\tilde F_B\biggl\{\frac{[k_1^2+k_2^2]}{\Lambda_B^2}\biggr\}
\nonumber
\end{eqnarray}
$\Gamma_i$ and $\lambda_{B_Q}$ are spin and flavor matrices,
respectively; $g_{B_Q}$ denotes the coupling of the heavy baryon with
the constituent quarks; $\Lambda_B$ is the cutoff
parameter defining the distributions of light quarks in the heavy baryon.
As we are working to leading order in the heavy quark mass expansion
the baryon cutoff parameter $\Lambda_B$ has to be chosen to be
the same for the charm and bottom baryons in order to guarantee the correct
normalization of the baryonic Isgur-Wise function in the heavy quark symmetry
limit~\cite{PRD97}. The quantum numbers of the heavy baryons and the Dirac
matrices $\Gamma_i$ and flavour matrices $\lambda_{B_Q}$ define the structure
of the relevant three-quark charm baryon currents. They are listed in TABLE I.

Let us now specify how the electromagnetic interactions
are introduced at the quark level.
As in a local theory we derive the interaction Lagrangian of the
electromagnetic field $A_\mu$ with the quarks using the standard minimal
substitution
\begin{eqnarray}
L_{em}(x)= - eA_\mu(x)\bar q^a(x)\gamma^\mu Q q^a(x)
\end{eqnarray}
where $Q=\mbox{diag}\{2/3, -1/3, -1/3\}$ is the charge matrix of the light
quarks. In the leading order of the heavy quark mass expansion the photon
does not couple to the heavy quark.
In order to have a gauge-invariant theory one also needs to couple the
electromagnetic field into the nonlocal heavy-baryon-quark Lagrangian
(\ref{HB_Lagr}). This can be achieved by the prescription of Mandelstam
\cite{Mandelstam}. Previous applications of the Mandelstam prescription to
hadron physics can be found  in Refs.~\cite{Terning,FBS}.
Each light quark field $q(y)$ is multiplied with the exponential
factor
$\exp(ieQ\int\limits_x^y dz_\mu A^\mu(z))$. As a result one obtains
a nonlocal gauge invariant interaction Lagrangian for the coupling
of heavy baryons to their constituent quarks including electromagnetism.
One has
\begin{eqnarray}\label{Lagr_GI}
{\cal L}_{B_Q}^{\rm int, e.m.}(x)&=&g_{B_Q}\bar B_Q(x)\Gamma_1 Q^a(x)
\int d^4\xi_1\int d^4\xi_2F_B(\xi_1^2+\xi_2^2)\\
&\times&\exp(ieQ\int\limits_x^{x+3\xi_1-\sqrt{3}\xi_2} dz_\mu A^\mu(z))
q^b(x+3\xi_1-\sqrt{3}\xi_2 )C\Gamma_2\lambda_{B_Q}\nonumber\\
&\times&\exp(ieQ\int\limits_x^{x+3\xi_1+\sqrt{3}\xi_2} dz_\mu A^\mu(z))
q^c(x+3\xi_1+\sqrt{3}\xi_2)\varepsilon^{abc}+{\rm h.c.}\nonumber
\end{eqnarray}
The Lagrangian (\ref{Lagr_GI}) generates nonlocal vertices which involve
the heavy baryons, photons and light and heavy  quarks. In general
several diagrams
contribute to the one-photon transitions of heavy baryons:
the standard triangle diagram (FIG.Ia) and
the contact interaction-type diagrams (FIG.Ib). The calculation of the
contact interaction-type
diagrams was discussed in detail in Ref. \cite{FBS} where this
approach was applied to the study of nucleon electro-magnetic interactions.
Only when the contact interaction-type diagrams are included one satisfies
the relevant Ward-Takahashi
identities for the connected Green functions (see details in Ref. \cite{FBS}).
However, it is not difficult to see that
the contact interaction-type contributions are nonleading in the heavy
mass expansion, at least when the photon is on its mass shell $(q^2=0)$.
Since we are working in the heavy quark limit and with real photons
throughout, these contributions
can be safely dropped. For transitions involving P-wave states there are,
however, leading contact interaction-type contributions resulting from the
minimal substitution prescription for the derivatives in the interaction
Lagrangian coupling the excited baryon states to the constituent quarks.

The contribution of the triangle diagram (FIG.Ia) to the matrix element of
the one-photon transition $B^i_Q(p)\to B^f_Q(p)+\gamma(q)$
has the following form in the heavy quark limit
\begin{eqnarray}\label{vertex}
M_{inv, \Delta}^\gamma(B_Q^i\to B_Q^f\gamma)=eg^i_{\rm eff} g^f_{\rm eff}
\cdot \bar u(v)\Gamma_1^f \frac{(1+\not\! v)}{2}\Gamma_1^iu(v)
\cdot I_{q_1q_2}^{if}(v,q)
\end{eqnarray}
$$
g_{\rm eff}=g_{B_Q} \frac{\Lambda_B^2 \sqrt{C_{\rm color}}}{8\pi^2},
\hspace{1.5cm} C_{\rm color}=6
$$
\begin{eqnarray}\label{int}
\hspace*{-1cm}I_{q_1q_2}^{if}(v,q)&=&
\int\hspace*{-0.1cm}\frac{d^4k_1}{\pi^2i} \hspace*{-0.1cm}
\int\hspace*{-0.1cm}\frac{d^4k_2}{4\pi^2i} \hspace*{0.1cm}
\frac{\tilde {\cal F_B}(k_1,k_2,q)\tilde {\cal F_B}(k_1,k_2,0)}
{[-k_1v-\bar\Lambda_{q_1q_2}]}
\Pi_{q_1q_2}(k_1,k_2,q)\\[2mm]
\tilde {\cal F_B}(k_1,k_2,q)&\equiv&
\tilde F_B\biggl\{-6\biggl[(k_1+q)^2+(k_2-q)^2+(k_1+k_2)^2\biggr]\biggr\}
\nonumber\\[2mm]
\Pi_{q_1q_2}(k_1,k_2,q)&=&
Q_{q_2q_2}{\rm tr}\biggl[\Gamma_2^i S_{q_1}(k_1+k_2)
\Gamma_2^fS_{q_2}(k_2-q)\gamma^\mu S_{q_2}(k_2)\biggr] \nonumber\\
&-&Q_{q_1q_1}{\rm tr}\biggl[\Gamma_2^f S_{q_2}(-k_1-k_2)\Gamma_2^i
S_{q_1}(-k_2)\gamma^\mu S_{q_1}(-k_2+q)\biggr]
\nonumber
\end{eqnarray}
where $\Gamma_{1(2)}^i$ and $\Gamma_{1(2)}^f$ are the
Dirac matrices of the initial and the final baryons, respectively.
Here $S_q(k)=1/(m_q - \not\!\!k)$ is the light quark propagator $(q=u,d,s)$.
The masses of the $u$ and the $d$ quarks are set equal: $m_u=m_d=m_q$.
The parameter $\bar\Lambda_{q_1q_2}=M_{Qq_1q_2}-m_Q$ in the denominator
of the heavy quark propagator denotes the difference
between the heavy baryon mass $M_{Qq_1q_2}$ and the heavy quark mass $m_Q$.
We use different values for the parameter $\bar\Lambda_{q_1q_2}$
for baryons containing only nonstrange light quarks and one or two strange
quarks: $\bar\Lambda$, $\bar\Lambda_s$ and $\bar\Lambda_{ss}$, respectively.
The appearance of unphysical imaginary
parts in the Feynman diagrams is avoided by imposing the condition that
the baryon mass is less than the sum of constituent quark masses.
In the case of heavy-light baryons this restriction implies that the
parameter $\bar\Lambda_{q_1q_2}$ must be less than the sum of light
quark masses. Latter constraint serves as an upper limit for our choices
of the parameter $\bar\Lambda_{q_1q_2}$.
All dimensional parameters are expressed in units of $\Lambda_B$.
The integrals are calculated in the Euclidean region both
for internal and external momenta. Finally, the results for the physical
region are obtained by analytic continuation of the external momenta after
the internal momenta have been integrated out.

In the calculation of (\ref{int}) we use the $\alpha$-parametrization for
quark propagators and the Laplace transform for the vertex function
\begin{eqnarray}\label{trick}
& &\frac{1}{A}=\int\limits_0^\infty\hspace*{-0.1cm}d\alpha e^{-\alpha A},
\hspace{1.5cm}
\tilde F_B(6X)=\int\limits_0^\infty\hspace*{-0.1cm}ds \tilde F^L_B(6s)e^{-sX}
\end{eqnarray}
The use of the Laplace transform allows one to perform the calculation of the
transition matrix elements for any given function $\tilde F_B$. In the
numerical analysis of one-photon transitions of heavy baryons
we will use a Gaussian vertex functions for heavy baryons in Eq.~(\ref{int}).
As an illustration of our calculational procedure we evaluate a typical
matrix element as e.g.
\begin{eqnarray}
R^{if}_{q_1q_2}(v,q)&=&
\int\hspace*{-0.1cm}\frac{d^4k_1}{\pi^2i} \hspace*{-0.1cm}
\int\hspace*{-0.1cm}\frac{d^4k_2}{\pi^2i}
\hspace*{0.1cm}\frac{\tilde {\cal F_B}(k_1,k_2,q)\tilde {\cal F_B}(k_1,k_2,0)}
{[-k_1v-\bar\Lambda_{q_1q_2}]}\\
&\times&{\rm tr}[\Gamma_2^iS_{q_2}(q_1+q_2)\Gamma_2^fS_{q_1}(k_2-q)\gamma^\mu
S_{q_1}(k_2)]\nonumber
\end{eqnarray}
Using the representation (\ref{trick}) we obtain
\begin{eqnarray}
\hspace*{-1cm}& &R^{if}_{q_1q_2}(v,q)=
\int\limits_0^\infty \hspace*{-0.1cm}ds_1 \tilde F_B^L(6s_1)
\int\limits_0^\infty \hspace*{-0.1cm}ds_2 \tilde F_B^L(6s_2)e^{2s_2q^2}
\int\limits_0^\infty \hspace*{-0.1cm}d^4 \alpha
e^{\alpha_3\bar\Lambda-(\alpha_1+\alpha_4)m_{q_1}^2-\alpha_2m_{q_2}^2}
\nonumber\\
\hspace*{-1cm}&\times&{\rm tr}\biggl[\Gamma_2^i
\biggl(m_{q_2}-\frac{\not\!\partial_1+\not\!\partial_2}{2}\biggr)\Gamma_2^f
\biggl(m_{q_1}-\frac{\not\!\partial_2}{2}-\not\! q\biggr)\gamma^\mu
\biggl(m_{q_1}-\frac{\not\!\partial_2}{2}\biggr)\biggr]\nonumber
\int\hspace*{-0.1cm}\frac{d^4k_1}{\pi^2i} \hspace*{-0.1cm}
\int\hspace*{-0.1cm}\frac{d^4k_2}{\pi^2i} \hspace*{0.1cm}
e^{kAk-2kB}\nonumber
\end{eqnarray}
The integration over $k_1$ and $k_2$ results in
\begin{eqnarray}
\hspace*{-1cm}& &R^{if}_{q_1q_2}(v,q)=
\int\limits_0^\infty \hspace*{-0.1cm}ds_1 \tilde F_B^L(6s_1)
\int\limits_0^\infty \hspace*{-0.1cm}ds_2 \tilde F_B^L(6s_2)e^{2s_2q^2}
\int\limits_0^\infty \hspace*{-0.1cm}d^3\alpha
e^{\alpha_3\bar\Lambda-(\alpha_1+\alpha_4)m_{q_1}^2-\alpha_2m_{q_2}^2}
\nonumber\\
\hspace*{-1cm}&\times&{\rm tr}\biggl[\Gamma_2^i
\biggl(m_{q_2}-\frac{\not\!\partial_1+\not\!\partial_2}{2}\biggr)\Gamma_2^f
\biggl(m_{q_1}-\frac{\not\!\partial_2}{2}-\not\! q\biggr)\gamma^\mu
\biggl(m_{q_1}-\frac{\not\!\partial_2}{2}\biggr)\biggr]
\frac{e^{-BA^{-1}B}}{|A|^2}
\nonumber
\end{eqnarray}
where
\begin{eqnarray}
k A k - 2 k B = \sum\limits_{i,j=1}^{2} k_i A_{ij} k_j -
2\sum\limits_{i=1}^{2} k_i B_i, \hspace*{1cm}
\not\!\partial_i = \frac{\partial}{\partial\not\!\!B_i}\nonumber
\end{eqnarray}
\[A_{ij}=\left(
\begin{array}{ll}
 \mbox{$2(s_1+s_2)+\alpha_2$}  &  \hspace*{.5cm}  \mbox{$s_1+s_2+\alpha_2$}\\
 \mbox{$s_1+s_2+\alpha_2$}     &  \hspace*{.5cm}
\mbox{$2(s_1+s_2)+\alpha_1+\alpha_2+\alpha_4$}
\end{array}
\right) \]
$$
B_1=-s_2q-\alpha_3v/2 \hspace{1cm}  B_2=(s_2+\alpha_1)q
$$
The kinematics of the one-photon transitions allows one to make use of the
approximation: $qv=(m_i^2-m_f^2)/(2m_i) \approx 0$ where $m_i$
and $m_f$ are the masses of the initial and the final baryons, respectively,
divided by $\Lambda_B$.
Then, by making the variable replacement $\alpha_i\to(s_1+s_2)\alpha_i$
and by using $\Gamma_2^i=\gamma_\nu$ and $\Gamma_2^f=\gamma_5$ the overlap
integral can be seen to be proportional to $q^\mu$ such that
$$R^{\mu\nu}_{q_1q_2}(v,q)=
4i\varepsilon^{\mu\nu\alpha\beta}q^\alpha v^\beta J,\,\,\,\,
 J=\int\limits_0^\infty\hspace*{-0.1cm}
\frac{d^3\alpha\alpha_1\alpha_3}{2|A|^2}
\tilde F_B^2(6z)\{m_{q_1}(A_{11}^{-1}+A_{12}^{-1})-m_{q_2}A_{12}^{-1}\}
$$
\[A_{ij}=\left(
\begin{array}{ll}
 \mbox{$2+\alpha_2$}      &    \hspace*{.2cm}    \mbox{$ 1+\alpha_2$}\\
 \mbox{$ 1+\alpha_2$}     &    \hspace*{.2cm}    \mbox{$2+\alpha_1+\alpha_2$}
\end{array}
\right),
\hspace*{.5cm}
A^{-1}_{ij}=\frac{1}{|A|}\left(
\begin{array}{ll}
 \mbox{$2+\alpha_1+\alpha_2$} &    \hspace*{.2cm}    \mbox{$-(1+\alpha_2)$}\\
 \mbox{$-(1+\alpha_2)$}       &    \hspace*{.2cm}    \mbox{$2+\alpha_2$}
\end{array}
\right) \]
The evaluation of the other remaining matrix elements proceeds along similar
lines.

At this point it is convenient to define a standard set of gauge invariant
coupling constants for the one-photon transitions discussed in this paper.
The general expansion of the transition matrix elements into
a minimal set of  gauge invariant covariants reads
\begin{eqnarray}\label{matrix_el}
M_{inv}^\gamma(\Sigma_c\to\Lambda_c\gamma) &=&
i\frac{2}{\sqrt{3}}f_{\Sigma_c\Lambda_c\gamma}
\bar u(v) \not\! q \not\!\varepsilon^*(q) u(v)\nonumber \\
M_{inv}^\gamma(\Sigma_c^*\to\Lambda_c\gamma) &=&
2f_{\Sigma_c^*\Lambda_c\gamma}
\bar u(v)\epsilon(\mu^* \varepsilon^* v k) u^\mu(v)\nonumber \\
M_{inv}^\gamma(\Lambda_{c1}\to\Lambda_c\gamma) &=&
\bar u(v)[F_{\Lambda_{c1}\Lambda_c\gamma} \cdot g^{\alpha\mu} vq +
G_{\Lambda_{c1}\Lambda_c\gamma}\cdot v^{\alpha}q^{\mu}]
\frac{\gamma^\mu\gamma^5}{\sqrt 3}u(v)\varepsilon^*_\alpha(q)
\nonumber\\
M_{inv}^\gamma(\Lambda_{c1}^*\to\Lambda_c\gamma) &=&
\bar u(v)[F_{\Lambda_{c1}\Lambda_c\gamma}^* \cdot g^{\alpha\mu} vq +
G_{\Lambda_{c1}\Lambda_c\gamma}^* \cdot v^{\alpha}q^{\mu}]
u^\mu(v)\varepsilon^*_\alpha(q)
\nonumber
\end{eqnarray}
In the heavy quark limit three of the coupling constants become related.
The relations read \cite{Korner,Cho}
$$f_{\Sigma_c\Lambda_c\gamma}=f_{\Sigma^*_c\Lambda_c\gamma}=f$$
$$F_{\Lambda_{c1}\Lambda_c\gamma}=F_{\Lambda^{*}_{c1}\Lambda_c\gamma}=F$$
$$G_{\Lambda_{c1}\Lambda_c\gamma}=G_{\Lambda^{*}_{c1}\Lambda_c\gamma}=G$$

Returning to our model calculation the coupling constant $f$ can be
represented as
\begin{eqnarray}
f&=&(\mu_1-\mu_2)
\frac{R_{\Sigma_Q\Lambda_Q\gamma}}{\sqrt{R_{\Lambda_Q}}\sqrt{R_{\Sigma_Q}}}
\\[2mm]
R_{\Sigma_Q\Lambda_Q\gamma}&=&\frac{1}{4}
\int\limits_0^\infty\hspace*{-.1cm}d^3\alpha\alpha_3(\alpha_1+\alpha_2)
\tilde F^2_B(6z)\frac{A_{11}^{-1}}{|A|^2}
\nonumber\\[2mm]
R_{B_Q}&=&
\int\limits_0^\infty\hspace*{-.1cm}d^3\alpha\alpha_3
\frac{\tilde F^2_B(6z)}{|A|^2}
\biggl\{1+d_{B_Q}\frac{\alpha_3}{m^2_q}\frac{\partial z}{\partial\alpha_3}
-\frac{\alpha_3^2}{4m^2_q}A_{12}^{-1}(A_{11}^{-1}+A_{12}^{-1})\biggr\}
\nonumber
\end{eqnarray}
where $\mu_i=e_i/(2m_q)$ is the magnetic moment of the i-th light quark.
Here
$$
z=\frac{\alpha_3^2}{4}A_{11}^{-1}+m^2_q(\alpha_1+\alpha_2)
-\bar\Lambda\alpha_3,
\hspace*{1cm}
d_{B_Q}=\left\{
\begin{array}{ll}
1 & \,\,\, \mbox{for} \,\,\, B_Q=\Lambda_{Q}\\
\frac{1}{2} & \,\,\, \mbox{for} \,\,\, B_Q=\Sigma_{Q}\\
\end{array}
\right.
$$
The calculation of the other two coupling factors $F$ and $G$ proceeds
along similar lines.

The one-photon decay rates can then be calculated
by using the general rate formula
\begin{eqnarray}\label{rate}
\Gamma = \frac{1}{2J+1}
\quad \frac{ \mid {\bf q} \mid}{8 \pi M_{B_Q}^{2}}
\sum_{spins} \mid M^{\gamma}_{inv} \mid^{2}
\end{eqnarray}
where ${\mid {\bf q} \mid} = qv =
(m_i^2-m_f^2)/(2m_i)$ is the photon momentum
in the rest frame of the decaying baryon. In terms of the above coupling
constants one obtains
\begin{eqnarray}\label{all_rates}
\Gamma\left(\Sigma_c\rightarrow \Lambda_c \gamma \right)&=&
\frac{4}{3\pi} f^2 {\mid {\bf q} \mid}^3 \frac{M_{\Lambda_c}}{M_{\Sigma_c}}
\nonumber\\
\Gamma\left(\Sigma^*_c\rightarrow \Lambda_c\gamma\right)&=&
\frac{4}{3\pi} f^2 {\mid {\bf q} \mid}^3 \frac{M_{\Lambda_c}}{M_{\Sigma_c^*}}
\\
\Gamma\left(\Lambda_{c1}\rightarrow\Lambda_c\gamma \right)&=&
\frac{1}{3\pi} \frac{3F^2-G^2}{2} {\mid {\bf q} \mid}^3
\frac{M_{\Lambda_c}}{M_{\Lambda_{c1}}}
\nonumber\\
\Gamma\left(\Lambda_{c1}^*\rightarrow\Lambda_c\gamma \right)&=&
\frac{1}{3\pi} \frac{3F^2-G^2}{2} {\mid {\bf q} \mid}^3
\frac{M_{\Lambda_c}}{M_{\Lambda_{c1}^*}}
\nonumber
\end{eqnarray}

Let us now specify our model parameters. In our numerical evaluation of
the one-photon transition rates we make use of the same set of model
parameters are were used to study the properties of light and heavy
baryons \cite{FBS,PRD97} and one-pion transitions between
charmed baryons \cite{MZ98}. In particular, the coupling constants
$g_{B_Q}$ in Eqs. (\ref{Lagr_GI}) are calculated
from {\it the compositeness condition} (see, ref.~\cite{PRD97}), which means
that the renormalization constant of the hadron wave function is set equal to
zero $Z_{B_Q}=1-g_H^2\Sigma^\prime_{B_Q}(M_{B_Q})=0$ where $\Sigma_{B_Q}$
is the heavy baryon mass operator.
The masses of the light non-strange $u$ and the $d$ quarks
($m_u=m_d=m_q$) were determined from an analysis of
nucleon data: $m_q$=420 MeV \cite{FBS}.
The parameters $\Lambda_{B}$, $m_s$, $\bar\Lambda$ are
taken from the analysis of the $\Lambda^+_c\to\Lambda^0+e^+ +\nu_e$ decay
data \cite{MZ98}. To reproduce the present average value of
$B(\Lambda_c^+\to\Lambda e^+ \nu_e$) = 2.2 $\%$ we
used the following values for our parameters:
$\Lambda_{B}$=1.8 GeV, $m_s$=570 MeV and $\bar\Lambda$=600 MeV.
The values of the unknown parameters $\bar\Lambda_s$ and $\bar\Lambda_{ss}$
were determined \cite{MZ98} from the relations
$\bar\Lambda_s = \bar\Lambda + (m_s - m)$
and $\bar\Lambda_{ss} = \bar\Lambda + 2(m_s - m)$, which give
$\bar\Lambda_s$ = 750 MeV and $\bar\Lambda_{ss}$ = 900 MeV. Using the values
of $\Lambda_{B}$=1.8 GeV and $\bar\Lambda$=600 MeV one obtains a
satisfactory fit to
the decay $\Lambda_b^0\to\Lambda_c^+ e^- \bar\nu_e$ decay: the width
$\Gamma(\Lambda_b^0\to\Lambda_c^+e^-\bar\nu_e)=5.4\times 10^{10}s^{-1}$
and the slope of the $\Lambda_b$ Isgur-Wise function $\rho^2 = 1.4$.
Hence, in this paper the model parameters are set to
$m_q$ = 420 MeV, $m_s$ = 570 MeV, $\Lambda_{B}$ = 1.8 GeV,
$\bar\Lambda$ = 600 MeV,
$\bar\Lambda_s$ = 750 MeV,
$\bar\Lambda_{ss}$ = 900 MeV.
 Finally, the mass values of the charm baryon states including current
experimental errors are taken from \cite{PDG,CLEO_98} (see TABLE I).
The masses of the excited bottom baryons $\Lambda_{b1}$ and $\Lambda_{b1}^*$
are estimated from the heuristic relation:
$m_{\Lambda_{b1}^{(*)}} = m_{\Lambda_{c1}^{(*)}} +
(m_{\Lambda_{b}^{0}} - m_{\Lambda_{c}^{+}})$.

We now present our numerical results for the one-photon decay rates
of heavy baryons. Our results are listed in TABLE II.
The errors in our rate values reflect the experimental errors in the charm
baryon masses \cite{PDG,CLEO_98} (see TABLE I).
For the sake of comparison we also list the results of the model
calculations \cite{Cho}-\cite{Chow} mentioned earlier on. Our results are
quite close to the results of the nonrelativistic quark model \cite{Cheng}.
In \cite{Cho} the coupling strengths were parametrized in terms of unknown
effective coupling parameters $c_{RT}$. A first rough estimate of the
unknown coupling parameters can be obtained by setting them equal to 1 on
dimensional grounds \cite{Cho}. As is evident from TABLE II such an estimate
is basically supported by our dynamical calculation. We do not agree with
the predictions on the charm and bottom p-wave decay rates of \cite{Chow}
except for the $\Lambda^*_{b1}\to \Lambda_b^0\gamma$ rate where we are closer
to the rate calculated in \cite{Chow}.

Recently the radiative decays of
bottom baryons were studied with the use of the light-cone QCD sum rules
\cite{Zhu} in the leading order of heavy quark effective theory.
For the decay rates of the $\Sigma_b$ and $\Sigma_b^*$
baryons to $\Lambda_b^0\gamma$ the authors of \cite{Zhu} obtained
$$
\Gamma(\Sigma_b\to\Lambda_b\gamma)=\alpha_{eff}  {\mid {\bf q} \mid}^3
\,\,\,\,\,
\mbox{and}\,\,\,\,\,
\Gamma(\Sigma_b^*\to\Lambda_b\gamma)=\alpha^*_{eff}  {\mid {\bf q} \mid}^3
$$
where the couplings $\alpha_{eff}$ and $\alpha^*_{eff}$ are approximately
equal to each other. The authors of \cite{Zhu} quote
$\alpha_{eff}\approx\alpha^*_{eff}\approx 0.03 $ GeV~$^{-2}$. In order to
compare our model results with the results in \cite{Zhu} we set
$M_{\Lambda_Q}=M_{\Sigma_Q}$ in Eq. (\ref{all_rates}). We then obtain
$\alpha_{eff} = 4f^2/(3\pi) \approx 0.015 $GeV$^{-2}$ which is one-half
the prediction of Ref.~\cite{Zhu}.

In conclusion, we have investigated electromagnetic decays of
heavy baryons. We have obtained predictions for the rates of the
two-body transitions $B^i_Q(p)\to B^f_Q(p^\prime)+\gamma(q)$.
We have compared our results with the results of other model
calculations~\cite{Cho}-\cite{Chow}. Unfortunately there is no data yet
to compare our results with. For the one-photon decays from
the $p$-wave states $\Lambda_{c1}\to\Lambda_c+\gamma$ and
$\Lambda_{c1}^*\to\Lambda_c+\gamma$ our predicted rates are one order of
magnitude below the upper limits given by the experiments calling for an
one-order of magnitude improvement of the experimental upper limits.
Although the $\Xi_c^\prime \to\Xi_c + \gamma$ one-photon decays have now been
seen \cite{CLEO_98} it will be close to impossible to obtain rate values for
these decays because the $\Xi_c^\prime$-states are far too narrow. The total
widths of the $\Sigma_c$, $\Sigma_c^*$ and $\Xi_c^*$ states are larger
because they also decay via one-pion emission. In fact the widths of the
$\Sigma_c^{*++}$ and $\Sigma_c^{*0}$ have been determined \cite{PDG}. One
can hope that one-pion branching ratios can be experimentally determined for
the $\Sigma_c$, $\Sigma_c^*$ and $\Xi_c^*$ one-photon decay modes in the
near future. We are looking forward to compare the predictions of
the Relativistic Three-Quark Model for the one-photon rates with
future experimental data.

\noindent
{\bf Acknowledgments}

\noindent
M.A.I and V.E.L thank Mainz University for the hospitality
where a part of this work was completed. This work was supported
in part by the Heisenberg-Landau Program, by the Russian Fund for
Basic Research (RFBR) under contracts 96-02-17435-a, 99-02-17731-a
and by the BMBF (Germany) under contract 06MZ865. J.G.K. acknowledges
partial support by the BMBF (Germany) under contract 06MZ865.
V.E.L. thanks the Russian Federal Program "Integration of
Education and Fundamental Science" for partial support.

\newpage
\centerline{\bf List of Figures}

\vspace*{.5cm}
\unitlength=1.00mm
\special{em:linewidth 0.4pt}
\linethickness{0.4pt}
\begin{picture}(115.00,64.00)
\put(50.00,20.00){\circle*{5.20}}
\put(95.00,20.00){\circle*{5.20}}
\put(73.00,42.00){\circle*{4.00}}
\put(73.00,5.00){\makebox(0,0)[cc]{\Large$\bf Q$}}
\put(30.00,26.00){\makebox(0,0)[cc]{\Large$\bf B_Q^i$}}
\put(115.00,26.00){\makebox(0,0)[cc]{\Large$\bf B_Q^f$}}
\put(73.00,23.20){\makebox(0,0)[cc]{\Large$\bf q_2$}}
\put(59.00,34.20){\makebox(0,0)[cc]{\Large$\bf q_1$}}
\put(87.00,34.20){\makebox(0,0)[cc]{\Large$\bf q_1$}}
\put(140.00,20.00){\makebox(0,0)[cc]
{\Large$\bf + \,\,\,(q_1\leftrightarrow q_2)$}}
\put(72.50,17.00){\oval(45.00,16.00)[b]}
\put(50.00,20.00){\line(1,1){23.00}}
\put(30.00,20.00){\line(1,0){85.00}}
\put(30.00,19.00){\line(1,0){20.00}}
\put(30.00,21.00){\line(1,0){20.00}}
\put(95.00,20.00){\line(0,-1){1.00}}
\put(95.00,19.00){\line(1,0){20.00}}
\put(95.00,21.00){\line(1,0){20.00}}
\put(95.00,20.00){\line(-1,1){23.00}}
\put(72.85,40.00){\line(0,1){2.00}}
\put(72.85,43.00){\line(0,1){2.00}}
\put(72.85,46.00){\line(0,1){2.00}}
\put(72.85,49.00){\line(0,1){2.00}}
\put(72.85,52.00){\line(0,1){2.00}}
\put(72.85,55.00){\line(0,1){2.00}}
\put(72.85,58.00){\line(0,1){2.00}}
\put(72.85,61.00){\line(0,1){2.00}}
\put(72,67){\makebox(0,0)[cc]{\LARGE$\bf \gamma$}}
\end{picture}

\centerline{\bf (a) Triangle diagram}

\vspace*{.75cm}
\unitlength=1.00mm
\special{em:linewidth 0.4pt}
\linethickness{0.4pt}
\begin{picture}(115.00,51.00)
\put(50.00,20.00){\circle*{5.20}}
\put(95.00,20.00){\circle*{5.20}}
\put(73.00,8.00){\makebox(0,0)[cc]{\Large$\bf Q$}}
\put(30.00,26.00){\makebox(0,0)[cc]{\Large$\bf B_Q^i$}}
\put(115.00,26.00){\makebox(0,0)[cc]{\Large$\bf B_Q^f$}}
\put(73.00,23.00){\makebox(0,0)[cc]{\Large$\bf q_2$}}
\put(73.00,32.00){\makebox(0,0)[cc]{\Large$\bf q_1$}}
\put(72.50,20.00){\oval(45.00,16.00)}
\put(30.00,20.00){\line(1,0){85.00}}
\put(30.00,19.00){\line(1,0){20.00}}
\put(30.00,21.00){\line(1,0){20.00}}
\put(95.00,20.00){\line(0,-1){1.00}}
\put(95.00,19.00){\line(1,0){20.00}}
\put(95.00,21.00){\line(1,0){20.00}}
\put(50,20){\line(0,1){2.00}}
\put(50,23){\line(0,1){2.00}}
\put(50,26){\line(0,1){2.00}}
\put(50,29){\line(0,1){2.00}}
\put(50,32){\line(0,1){2.00}}
\put(50,35){\line(0,1){2.00}}
\put(50,38){\line(0,1){2.00}}
\put(50,41){\line(0,1){2.00}}
\put(50,50){\makebox(0,0)[cc]{\LARGE$\bf \gamma$}}
\end{picture}

\vspace*{.3cm}
\unitlength=1.00mm
\special{em:linewidth 0.4pt}
\linethickness{0.4pt}
\begin{picture}(115.00,51.00)
\put(50.00,20.00){\circle*{5.20}}
\put(95.00,20.00){\circle*{5.20}}
\put(73.00,8.00){\makebox(0,0)[cc]{\Large$\bf Q$}}
\put(30.00,26.00){\makebox(0,0)[cc]{\Large$\bf B_Q^i$}}
\put(115.00,26.00){\makebox(0,0)[cc]{\Large$\bf B_Q^f$}}
\put(73.00,23.00){\makebox(0,0)[cc]{\Large$\bf q_2$}}
\put(73.00,32.00){\makebox(0,0)[cc]{\Large$\bf q_1$}}
\put(72.50,20.00){\oval(45.00,16.00)}
\put(30.00,20.00){\line(1,0){85.00}}
\put(30.00,19.00){\line(1,0){20.00}}
\put(30.00,21.00){\line(1,0){20.00}}
\put(95.00,20.00){\line(0,-1){1.00}}
\put(95.00,19.00){\line(1,0){20.00}}
\put(95.00,21.00){\line(1,0){20.00}}
\put(95,20){\line(0,1){2.00}}
\put(95,23){\line(0,1){2.00}}
\put(95,26){\line(0,1){2.00}}
\put(95,29){\line(0,1){2.00}}
\put(95,32){\line(0,1){2.00}}
\put(95,35){\line(0,1){2.00}}
\put(95,38){\line(0,1){2.00}}
\put(95,41){\line(0,1){2.00}}
\put(95,50){\makebox(0,0)[cc]{\LARGE$\bf \gamma$}}
\end{picture}

\centerline{\bf (b) Contact interaction-type diagrams}

\vspace*{1cm}
\centerline{\bf FIG. I:
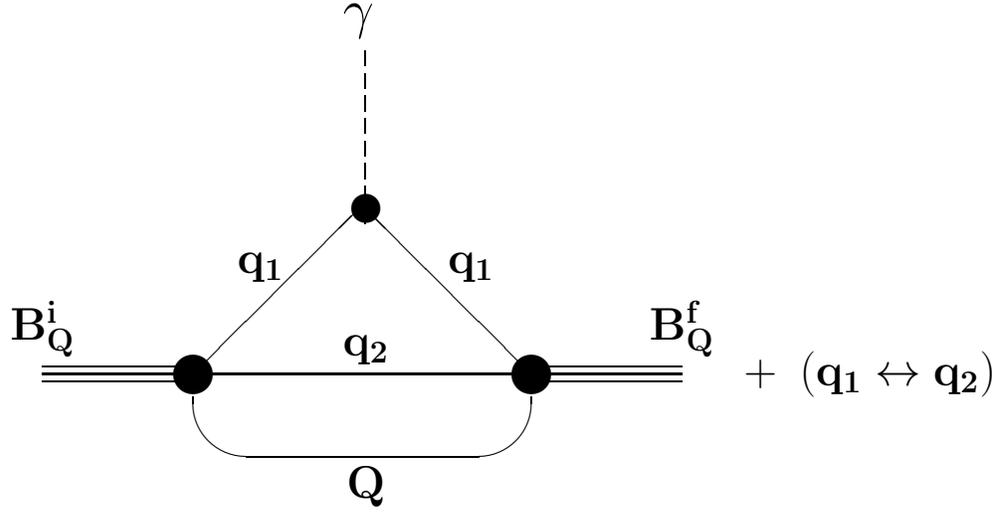
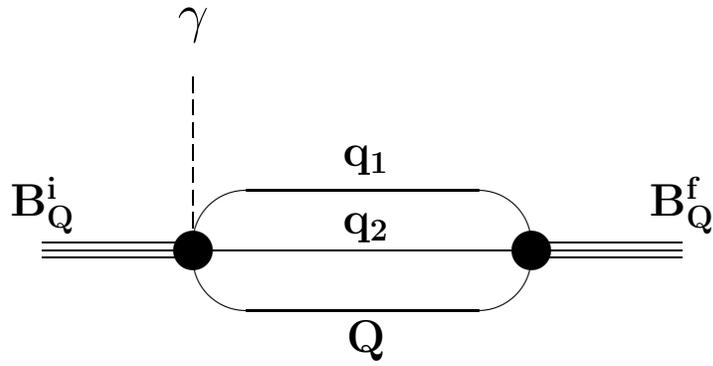
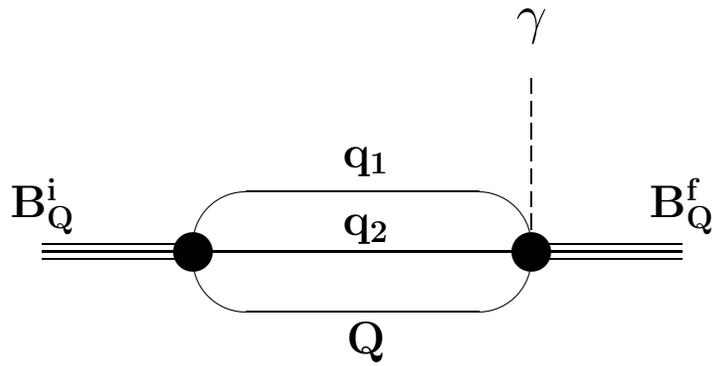
Diagrams contributing to one-photon heavy baryon transition
$B_Q^i\to B_Q^f\gamma$.}

\newpage
\centerline{\bf List of Tables}
\noindent
{\bf TABLE I}
Masses and spin and flavour quantum numbers of charm and bottom baryons.
Column 4 gives the structure of the coupling of the quark constituents
where
$\stackrel{\leftrightarrow}\partial_\mu = \stackrel{\leftarrow}\partial_\mu+
\stackrel{\rightarrow}\partial_\mu$.
(The square brackets $[...]$ and curly brackets $\{...\}$ denote
antisymmetric and symmetric flavour and spin combinations of
the light degrees of freedom. The $\lambda_i$ in column 5 are the usual
Gell-Mann matrices and $\lambda_u$ = diag\{1,0,0\},
$\lambda_d$ = diag\{0,1,0\}.

\vspace*{.2cm}
\noindent
{\bf TABLE II}
Decay rates $\Gamma$  for heavy baryon states.

\vspace*{3cm}
\begin{center}
{\bf TABLE I}
\end{center}
\def\arraystretch{1.2}
\begin{center}
\begin{tabular}{|c|c|c|c|c|c|}
\hline
Baryon&$\;J^P\;$&Quark & $\Gamma_1\otimes C\Gamma_2$
& $\; \lambda_{B_Q}\;$& $\;$Mass (MeV) \cite{PDG} $\;$\\
      & & Content & & &\\
\hline
$\Lambda_c^+$    & ${\frac{1}{2}}^+$&c[ud] &$I\otimes C\gamma^5$ & $i\lambda_2/2$ & $2284.9\pm 0.6$\\
\hline
$\Xi_c^+$        & ${\frac{1}{2}}^+$&c[us] &$I\otimes C\gamma^5$ & $i\lambda_5/2$ & $2465.6\pm 1.4$\\
\hline
$\Xi_c^0$        & ${\frac{1}{2}}^+$&c[ds] &$I\otimes C\gamma^5$ & $i\lambda_7/2$ & $2470.3\pm 1.8$\\
\hline
$\Xi_c^{+\prime}$& ${\frac{1}{2}}^+$&c\{us\} &$\gamma^\mu\gamma^5 \otimes C\gamma_\mu$ & $\lambda_4/(2\sqrt{3})$ & $2.5734\pm 3.1$\\
\hline
$\Xi_c^{0\prime}$& ${\frac{1}{2}}^+$&c\{ds\} &$\gamma^\mu\gamma^5 \otimes C\gamma_\mu$ & $\lambda_6/(2\sqrt{3})$ & $2.5773\pm 3.2$\\
\hline
$\Sigma_c^{++}$  & ${\frac{1}{2}}^+$&c\{uu\} &$\gamma^\mu\gamma^5 \otimes C\gamma_\mu$ & $\lambda_u/\sqrt{6}$ & $2452.8\pm 0.6$\\
\hline
$\Sigma_c^+$     & ${\frac{1}{2}}^+$&c\{ud\} &$\gamma^\mu\gamma^5 \otimes C\gamma_\mu$ & $\lambda_1/(2\sqrt{3})$ & $2453.6\pm 0.9$\\
\hline
$\Sigma_c^0$     & ${\frac{1}{2}}^+$&c\{dd\} &$\gamma^\mu\gamma^5 \otimes C\gamma_\mu$ & $\lambda_d/\sqrt{6}$ & $2452.2\pm 0.6$\\
\hline
$\Xi_c^{*+}$     & ${\frac{3}{2}}^+$&c\{us\} &$I \otimes C\gamma_\mu$ & $\lambda_4/2$ & $2644.6\pm 2.1$\\
\hline
$\Xi_c^{*0}$     & ${\frac{3}{2}}^+$&c\{ds\} &$I \otimes C\gamma_\mu$ & $\lambda_6/2$ & $2643.8\pm 1.8$\\
\hline
$\Sigma_c^{*++}$ & ${\frac{3}{2}}^+$&c\{uu\} &$I \otimes C\gamma_\mu$ & $\lambda_u/\sqrt{2}$ & $2519.4\pm 1.5$\\
\hline
$\Sigma_c^{*0}$    & ${\frac{3}{2}}^+$&c\{dd\} &$I \otimes C\gamma_\mu$ & $\lambda_d/\sqrt{2}$ & $2517.5\pm 1.4$\\
\hline
$\Lambda_{c1}$   & ${\frac{1}{2}}^-$  &c[ud]
&$\gamma^\mu\gamma^5\otimes C\gamma^5\stackrel{\leftrightarrow}\partial_\mu $
& $i\lambda_2/(2\sqrt{3})$ & $2593.9\pm 0.8$\\
\hline
$\Lambda_{c1}^*$ & ${\frac{3}{2}}^-$  &c[ud]  &
$I \otimes C\gamma^5\stackrel{\leftrightarrow}\partial_\mu $
& $i\lambda_2/2$ & $2626.6\pm 0.8$\\
\hline
$\Xi_{c1}^*$     & ${\frac{3}{2}}^-$  &c[us]
&$I \otimes C\gamma^5 \stackrel{\leftrightarrow}\partial_\mu $
& $i\lambda_5/2$ & $2815.0 \pm 2.1$ \\
\hline
$\Lambda_b^0$    & ${\frac{1}{2}}^+$&b[ud] &$I\otimes C\gamma^5$
& $i\lambda_2/2$ & $5624\pm 9$\\
\hline
$\Lambda_{b1}$    & ${\frac{1}{2}}^+$&b[ud]
&$\gamma^\mu\gamma^5\otimes C\gamma^5\stackrel{\leftrightarrow}\partial_\mu $
& $i\lambda_2/(2\sqrt{3})$ & $5933\pm 10$\\
\hline
$\Lambda_{b1}^*$  & ${\frac{1}{2}}^+$&b[ud]
&$I \otimes C\gamma^5 \stackrel{\leftrightarrow}\partial_\mu $
& $i\lambda_2/2$ & $5966\pm 10$\\
\hline
\end{tabular}
\end{center}

\newpage
\begin{center}
{\bf TABLE II}
\end{center}
\small \normalsize
\begin{center}
\begin{tabular}{|c|c|c|c|c|}
\hline
$B_Q\rightarrow B^{\prime}_{Q}\gamma$ & This approach & Other approaches &
Experiment \cite{PDG}\\
\hline
$\Sigma^{+}_{c}\rightarrow \Lambda_c^+\gamma$ & $ 60.7\pm 1.5$ KeV
& $93$ KeV \cite{Cheng} & $  $\\
\hline
$\Sigma^{*+}_{c}\rightarrow\Lambda_c^+\gamma$ & $ 151 \pm 4$ KeV& & \\
\hline
$\Xi^{'+}_{c}\rightarrow \Xi_c^+\gamma$  &$ 12.7\pm 1.5$ KeV
& $16$ KeV \cite{Cheng} &\\
\hline
$\Xi^{'0}_{c}\rightarrow\Xi_c^0\gamma$&$0.17\pm 0.02$ KeV
&$0.3$ KeV \cite{Cheng}&\\
\hline
$\Xi^{*+}_{c}\rightarrow\Xi_c^+\gamma$&$54\pm 3$ KeV & &\\
\hline
$\Xi^{*0}_{c}\rightarrow\Xi_c^0\gamma$&$0.68 \pm 0.04$ KeV  & & \\
\hline
$\Lambda_{c1}(2593) \rightarrow \Lambda_c^+\gamma $ & $0.104\pm 0.001$ MeV
& $0.191c^2_{RT}$ MeV \cite{Cho} & $< 2.36^{+1.31}_{-0.85}$ MeV \\
 & & $0.016$ MeV \cite{Chow} &\\
\hline
$\Lambda_{c1}^*(2625) \rightarrow \Lambda_c^+\gamma$&$0.137\pm 0.002$ MeV
& $0.253c^2_{RT}$ MeV \cite{Cho} & $< 1$ MeV\\
 & & $0.021$ MeV \cite{Chow} &\\
\hline
$\Xi_{c1}^{*+}(2815) \rightarrow \Xi_c^+\gamma $ & $0.177\pm 0.005$ MeV& &\\
\hline
$\Xi_{c1}^{*0}(2815) \rightarrow \Xi_c^0\gamma $ & $0.463\pm 0.014$ MeV& &\\
\hline
$\Lambda_{b1} \rightarrow \Lambda_b^0\gamma $ & $0.126\pm 0.022$ MeV
& $0.09$ MeV \cite{Chow}& \\
\hline
$\Lambda_{b1}^* \rightarrow \Lambda_b^0\gamma $ & $0.168\pm 0.026$ MeV
& $0.119$ MeV \cite{Chow}& \\
\hline
\end{tabular}
\renewcommand{\baselinestretch}{1}
\small \normalsize
\end{center}
\end{document}